\documentstyle[12pt,epsfig]{article}
\newcommand{\simlt}
{\mbox{\raisebox{-0.5ex}{$\textstyle \; \sim$}
\raisebox{ 0.8ex}{$\textstyle  \!\!\!\!\!\!\! <$  }}}
\newcommand{\simgt}
{\mbox{\raisebox{-0.5ex}{$\textstyle \; \sim$}
\raisebox{ 0.8ex}{$\textstyle  \!\!\!\!\!\!\! >$  }}}
\begin{document}
\addtocounter{footnote}{1}
\title{
Dark matter in the Galaxy
}
\author{Neven Bili\'{c}\thanks{Permanent
 address:
Rudjer Bo\v skovi\'c Institute,
P.O.\ Box 180, 10002 Zagreb, Croatia;
 \hspace*{5mm} Email: bilic@thphys.irb.hr}\ ,
Gary B.\ Tupper, and Raoul D.\ Viollier\thanks{
 Email: viollier@physci.uct.ac.za}
\\
Institute of Theoretical Physics and Astrophysics, \\
 Department of Physics, University of Cape Town,  \\
 Private Bag, Rondebosch 7701, South Africa \\
 }
\maketitle
\begin{abstract}
After a brief  introduction to
standard cosmology and the  dark matter problem
in the Universe,
we consider a self-gravitating noninteracting
fermion gas at nonzero temperature as a model for the
dark matter halo of the Galaxy.
This fermion gas model is
then shown to imply the existence of a supermassive
compact dark object at the Galactic center.
\end{abstract}

\section{Introduction}
At some stage of the evolution of the Universe,
primordial density fluctuations must have become gravitationally
unstable forming
dense clumps of dark matter (DM) that
 have survived until today
in the form of galactic halos.
In the recent past,
galactic halos were  successfully modeled as
a self-gravitating
isothermal gas of particles of arbitrary mass,
the density of which  scales asymptotically as
$r^{-2}$, yielding flat rotation curves
\cite{col}.
The aim of this paper is to describe the halo
of our Galaxy
in terms of a self-gravitating fermion gas
in hydrostatic and thermal equilibrium
at finite temperature.

Self-gravitating weakly interacting fermionic matter
has been exploited in a wide range of
astrophysical phenomena.
Originally,
self-gravitating degenerate
neutrino stars were suggested as a model
for quasars
\cite{1}, and later
neutrino matter was used
as a model for dark matter in galactic halos and
clusters of galaxies, with a neutrino mass in the $\sim$ eV
range \cite{cos}.
Recently,
degenerate superstars, composed of
weakly interacting fermions in the $\sim$ 10 keV range,
were suggested
\cite{2,3,4,5,6}
as an alternative to the supermassive black holes that are
believed to exist at the centers of galaxies.
It was shown \cite{4} that such degenerate fermion stars
could explain the whole range of supermassive compact dark objects
which have been observed so far, with masses ranging from $10^{6}$ to
$3 \times 10^{9}$$M_{\odot}$, merely assuming that a weakly
interacting quasistable fermion of mass $m \simeq$ 15 keV exists
in Nature.
Most recently, it has been
pointed out that a weakly
interacting dark matter particle in the mass
range $1 \simlt m/{\rm keV} \simlt 5$
could solve the problem of the excessive
structure generated on subgalactic scales in
$N$-body and hydrodynamical
simulations of structure formation in this Universe \cite{bode}.

Of course, it is well known that
the interval
1-15 keV lies squarely in the
cosmologically forbidden mass range for stable active neutrinos $\nu$
\cite{11}.
The existence of such a massive active neutrino is also
disfavored by the Super-Kamiokande data
\cite{fuk}.
However,
for an initial lepton asymmetry of $\sim
10^{-3}$, a sterile neutrino $\nu_{s}$ of mass $m_{s} \sim 10$ keV
may be resonantly produced in the early Universe
with near closure density, i.e., $\Omega \simeq 1$
\cite{shi}.
The resulting energy spectrum is not thermal
but rather cut off so that it
approximates a degenerate Fermi gas.
In this mass range, sterile neutrinos are also
constrained
by astrophysical bounds on
the radiative
$\nu_{s} \rightarrow \nu \gamma$ decay \cite{13}.
However, the allowed parameter space includes $m_{s} \simeq 15$ keV,
contributing $\Omega_{d} \simeq 0.3$
to the critical density,
as favored by the BOOMERANG data
\cite{14}.
As an alternative possibility,
the  $\sim$ 15 keV sterile neutrino
could be replaced by the axino
\cite{cov}
or the gravitino
\cite{din,mur}
in soft supersymmetry breaking scenarios.

As the supermassive compact
dark objects at the galactic centers
are well described by a degenerate gas of fermions,
it is tempting to explore the
possibility that one could describe
both the supermassive compact dark objects
and their galactic halos in
a unified way in terms of a fermion gas
at finite temperature.
We will show  that this
is indeed the case, and that
the observed dark matter distribution in the
Galactic halo is consistent with
the existence of a supermassive compact dark
object at the center of
the Galaxy which has about the right mass and size,
and is in thermal and hydrostatic equilibrium
with the halo.

\section{Standard cosmology}
Standard cosmology provides a successful
description of the evolution of the Universe from
a fraction of a second after the creation until today.
A short review  on the standard model of cosmology
is given in \cite{rev1}.
For our purpose, it is sufficient to state the basic
underlying principles.
Standard cosmology is based on the
following three theoretical assumptions:

{\bf 1.\ Cosmological principle}.
The cosmological principle asserts
that the Universe is homogeneous and isotropic on large scales.
The most general metric satisfying the cosmological principle
is the Friedmann-Robertson-Walker  metric \cite{wei}
\begin{equation}
ds^2= dt^2 -a(t)^2\left[\frac{dr^2}{1-kr^2} -
     r^2(d\theta^2+\sin \theta d\phi^2)\right],
\label{eq101}
\end{equation}
where the curvature constant $k$ takes on the values
1, 0, or -1, for a closed,
flat, or open universe, respectively.
The time-dependent quantity $a(t)$
is the scale factor of the expansion
conveniently normalized to unity at present time,
i.e., $a(t_0)=1$. In other words,
$a$ is the radius of the Universe
measured in units of its current radius.

{\bf 2.\ General relativity}.
Gravity is described by Einstein's general theory of relativity
governed by the {\em equivalence principle}
and Einstein's field equations.

{\bf 3.\ Perfect fluid}. Matter is approximated by a homogeneous
perfect fluid.
The energy-momentum tensor then takes
a simple form
\begin{equation}
T_{\mu\nu}=
(\rho+p)u_{\mu}u_{\nu}
-g_{\mu\nu}\, ,
\label{eq102}
\end{equation}
where $\rho$ and $p$ are the density and the pressure
of the fluid, respectively.

With these assumptions, the set of Einstein's
equations reduces to the
Friedmann-Robertson-Walker (FRW) equations
\begin{equation}
H(t)^2\equiv \left(\frac{\dot{a}}{a}\right)^2=
    \frac{8\pi\rho}{3}-\frac{k}{a^2}
    +\frac{\Lambda}{3} \, ,
\label{eq103}
\end{equation}
\begin{equation}
\frac{\ddot{a}}{a}=
    \frac{\Lambda}{3}
  -4\pi(\rho+3p),
\label{eq104}
\end{equation}
where the natural system of units
$\hbar=c=G=1$ is assumed.
The first FRW equation describes the expansion
of the Universe.
The quantity $H(t)$ is the Hubble ``constant''
and $\Lambda$ is the cosmological constant.
We define the critical density as
$\rho_{\rm cr}=3 H_0^2/8\pi$ and the ratio of the
density to the critical density is denoted by
$\Omega=\rho/\rho_{\rm cr}$.
The precise present value of the
 Hubble constant is not known,
 but the widely accepted value is
 $H_0=100 h_0\,{\rm km s^{-1} Mpc^{-1}}$,
with a dimensionless parameter $h_0$ between 0.4 and 1.
Dividing by $H(t_0)^2$,
Eq.\ (\ref{eq103}) at $t=t_0$
 can be conveniently written as a sum rule, i.e.,
\begin{equation}
\Omega_0
-k/(H_0a_0)^2+
\Omega_{\Lambda}=1,
\label{eq201}
\end{equation}
where
$\Omega_{\Lambda}=
\Lambda/(3 H_0^2)$
is the vacuum energy
contribution to the critical density
today.
Observational evidence favors  a flat universe
today, i.e., $k=0$.
Thus, Eq.\ (\ref{eq201}) becomes
\begin{equation}
\Omega_{\rm 0}+
\Omega_{\Lambda}=1.
\label{eq202}
\end{equation}
The second FRW equation (\ref{eq104}) describes the acceleration
of the expansion.
The expansion will accelerate or decelerate,
depending on whether
the vacuum energy dominates the matter
or vice versa.
However, even if
the cosmological term vanishes,
the expansion could accelerate if the dominant component
of the DM  obeys a peculiar equation of state
such that the pressure is negative and higher than
one third of the density.
One popular example is the scalar field
 model called {\em quintessence}
\cite{cal}.
Another scenario is based on a fluid obeying
the Chaplygin gas equation of state
$p \propto -1/\rho$ \cite{kam1,bil2},
which has been intensively
investigated for its solubility in 1+1-dimensional space-time,
for its supersymmetric
extension and connection to $d$-branes \cite{jac1}.

The observational evidence in support of
 standard cosmology may be summarized in
 four empirical pillars \cite{tur}:
Hubble's law, cosmic background radiation (CBR),
anisotropy of CBR, and abundance patterns of light elements.
However, despite an overwhelming observational
support, a number of problems remain
unsolved:
\begin{itemize}
\item
What caused the Big Bang and the expansion?
Did the Universe begin with more than 3+1 dimensions?
\item
Why is the cosmological constant $\Lambda$  about
50-120 orders of magnitude smaller than the value
expected from  quantum field theory?
\item
What caused the initial baryon-antibaryon
asymmetry that  led to the absence of antimatter
today?
\item
Why is the Universe so smooth on large scales
as evidenced by CBR?
\item
What caused the primordial density fluctuations that
provided the seeds  for structure formation?
\item
What does  nonbaryonic DM consist of?
\end{itemize}
A solution to these problems most probably goes beyond the
standard model of cosmology and certainly beyond
the standard model of particle physics.

\section{Dark matter}

Here, we briefly discuss the
DM problem and possible
DM candidates.
A more detailed analysis  may be found in
 a number of recent review articles
 \cite{dol}.

DM has to be introduced because
of the following facts:
\begin{itemize}
\item
Astronomical observations, such as the flattness
of the rotation curves of spiral galaxies and
the peculiar motion of galaxies within clusters,
strongly indicate that
\begin{equation}
\Omega_0\equiv \frac{\rho_{\rm matt}}{\rho_{\rm cr}}
\simgt 0.3 \, .
\label{eq105}
\end{equation}
\item
Consistency with the Big Bang nucleosynthesis
implies for  baryonic matter
\begin{equation}
0.008 h_0^{-2}
< \Omega_{B} <
0.024 h_0^{-2} .
\label{eq106}
\end{equation}
\item
Astronomical observations yield a small relative density of
luminous matter
\begin{equation}
 \Omega_{\rm lum}=
0.003 h_0^{-1}  .
\label{eq107}
\end{equation}
\end{itemize}

From these facts we conclude that
\begin{itemize}
\item
About 95\% of matter is {\bf dark}.
\item
About 60\%-90\%  of DM is {\bf nonbaryonic}.
\item
At least 75\% of baryonic matter is {\bf  dark}.
\end{itemize}
Whereas  baryonic DM is most likely
in the form of relatively standard astrophysical objects,
 e.g.,  cold hydrogen clouds or compact
objects such as neutron stars, brown dwarfs, MACHOs,
and even black holes,
the nature of nonbaryonic DM is unknown and
still a subject of  speculations.
Candidates within the standard model are practically
excluded
and those beyond the standard model have not yet been
detected in particle physics experiments.
Nevertheless, cosmological and astrophysical observations
tell us what these, yet undetected, particles could
or could not be.

The different DM scenarios are conveniently classified
as {\em hot, warm,} and {\em cold} DM
\cite{pri},
depending on the thermal velocities of DM particles in the
early Universe.

{\bf Hot} DM refers to low-mass neutral particles that are still
relativistic when galaxy-size masses ($\sim 10^{12}M_{\odot}$)
are first encompassed within the horizon.
Hence, fluctuations on galaxy scales are wiped out by the
``free streaming" of the  dark matter.
Standard examples of hot DM are
neutrinos and majorons.
They are still in thermal equilibrium after the QCD deconfinement
transition, which took place at $T_{\rm QCD}\simeq$ 150 MeV.
Hot DM particles have a cosmological number
density comparable with that of  microwave
background photons, which implies
an upper bound to their mass of a few tens of eV.

{\bf Warm} DM particles are just becoming nonrelativistic
when galaxy-size masses enter the horizon.
Warm DM particles interact much more weakly than neutrinos.
They decouple (i.e., their mean free path first exceeds
the horizon size) at $T \gg T_{\rm QCD}$.
As a consequence, their number is expected to be
roughly an order of magnitude lower and their mass an
order of magnitude larger, than hot DM particles.
Examples of warm DM are
 $\sim$ keV sterile neutrinos,
axinos
\cite{cov},
or gravitinos
in soft supersymmetry breaking scenarios
\cite{din,mur}.
There has been renewed interest in the
standard model neutrino as a candidate
for warm DM \cite{giu}.

{\bf Cold} DM particles are already nonrelativistic
when even globular cluster masses
($\sim 10^6M_{\odot}$) enter the horizon.
Hence, their free streaming  is of no cosmological
importance. In other words, all cosmologically
relevant fluctuations survive in a universe dominated
by cold DM.
The two main particle candidates for cold dark matter
are the lowest supersymmetric
{\em weakly interacting massive particles} (WIMPs)
and the {\em axion}.

One of the central issues in
 dark-matter  modeling
 is  the problem of
structure formation on subgalactic scales.
The combination of cold DM
and a small cosmological constant
($\Lambda$CDM)
seems to be in good agreement with many observational constraints.
However,
N-body and hydrodynamical simulations of galaxy formation
evidence that
$\Lambda$CDM overpredicts structure on small scales
\cite{bode}.
In addition to that,  high-resolution simulations
generally find a dark matter profile with a central
cusp $\rho \propto r^{-1.5}$ for galactic
halos \cite{kly,moo}
which seems to contradict the observations.

Clustering on small scales could be suppressed by an upper limit
to the phase-space density of DM particles
owing either to degeneracy pressure if they are fermions
or to a repulsive interaction if they are bosons.
The fermion mass would have to lie in the range
$0.1 \simlt m \simlt 10$ keV.
A specific scenario invoking keV mass fermions is
the {\em cool-dark-matter} proposal \cite{13}
for which candidates exist in shadow-world models and
the axino.

\section{Galactic halo}
We now discuss
the properties of the halo
of our Galaxy
assuming that it consists of a self-gravitating gas
of keV mass fermions
in hydrostatic and thermal equilibrium
at finite temperature.
The Milky Way Galaxy consists of five major components
which are nested within each other
\cite{bus}.
A spheroidal {\bf halo} with modest concentations of stars
and about 170 globular clusters extends out to a
radius of perhaps 200 kpc.
Within a radius of $\sim 25$ kpc, the halo contains
stars and open clusters that are concentrated into two
essentially coplanar disks:
 the {\bf thin disk} and the {\bf thick disk}.
In their innermost part the disks merge with a spheroidal {\bf bulge},
the central concentration of luminous matter in the
Galaxy.
Finally,
a compact dark object
 with a mass of $M_{\rm c} \simeq 2.6 \times
10^6 M_{\odot}$ is located
in the vicinity of the
enigmatic radio source Sgr A$^{*}$ at the
Galactic center
\cite{10},
within a radius of
18 mpc.

Here, we demonstrate that by extending
the Thomas-Fermi theory
to nonzero temperature
it is possible to
explain, within the same model, both the Galactic halo and the
compact dark object at the Galactic center.
Extending the Thomas-Fermi theory
to finite temperature \cite{17,16,bil}, it has been
shown that, at  some
critical temperature $T_{\rm c}$,
weakly interacting massive fermionic matter
with a total mass below the Oppenheimer-Volkoff limit
\cite{opp}
undergoes a first-order gravitational phase transition from a diffuse
to a clustered state, i.e., a nearly degenerate fermion star.
However, during this first-order phase transition
a  large amount of latent heat must be released in order
to substantially decrease
the entropy of the initial diffuse configuration.
In the absence of a  mechanism which would
make such a release possible,
the system would remain in a thermodynamic
quasistable supercooled state close to the point of gravothermal
collapse.
The Fermi gas will be caught in the supercooled
state even if the total mass of the gas exceeds the
Oppenheimer-Volkoff limit
as a
stable condensed state does not exist in this case.

The formation of a quasistable supercooled state
may be understood as a process similar
to that of violent relaxation,
which was introduced to describe rapid
virialization of stars of different mass in globular clusters
\cite{lynd,bin}.
Through the gravitational collapse of an
overdense fluctuation  at about one Gyr after the Big Bang,
 part of gravitational energy transforms into the kinetic energy
of random motion of small-scale density fluctuations.
The resulting virialized
cloud will thus be well
approximated by a gravitationally stable thermalized
halo.
In order to estimate the mass-to-temperature ratio,
 we assume that an overdense cloud
of mass
 $M$ stops  expanding at the time
 $t_{\rm{m}}$, reaching its maximal radius $R_{\rm{m}}$ and the
 average density $\rho_{\rm{m}}=3 M/(4\pi R_{\rm{m}}^3)$.
 The total energy per particle is just the gravitational energy
 \begin{equation}
 E=-\frac{3}{5}\frac{M}{R_{\rm{m}}}\, .
 \label{eq001}
 \end{equation}
 From the spherical model of  nonlinear collapse
 \cite{padma}
 it follows
 \begin{equation}
 \rho_{\rm{m}}=\frac{9\pi^2}{16} \bar{\rho}(t_{\rm{m}})
 =\frac{9\pi^2}{16} \Omega_d \rho_{\rm{cr}} (1+z_{\rm{m}})^3,
 \label{eq002}
 \end{equation}
 where $\bar{\rho}(t_{\rm{m}})$ is the background density
 at the time $t_{\rm{m}}$ or  the cosmological
 redshift $z_{\rm{m}}$.
We approximate the virialized cloud by
a singular isothermal sphere
\cite{bin}
of the
 mass of the Galaxy $M$ and
radius $R$.
The singular isothermal sphere is characterized by
a constant circular velocity
$ \Theta=(2 T/m)^{1/2}$
and the density profile
 \begin{equation}
 \rho(r)=\frac{\Theta^2}{4\pi  r^2}.
 \label{eq003}
 \end{equation}
Its total energy per particle is the sum of gravitational
and thermal energies, i.e.,
 \begin{equation}
 E=-\frac{1}{4}\frac{M}{R}
 =-\frac{1}{4}\Theta^2 .
 \label{eq004}
 \end{equation}
 Combining Eqs.\ (\ref{eq001}), (\ref{eq002}),
 and (\ref{eq004}),
 we find
 \begin{equation}
 \Theta^2=\frac{6\pi}{5}
 (6 \Omega_d \rho_{\rm{cr}} M^2)^{1/3}(1+z_{\rm{m}})  .
 \label{eq005}
 \end{equation}
 Taking $\Omega_{\rm{d}}=0.3$, $M=2\times 10^{12} M_{\odot}$,
 $z_{\rm{m}}=4$, and $h_0=0.65$, we find
 $\Theta \simeq 220\, {\rm km\, s^{-1}}$, which corresponds to the
 mass-temperature ratio $m/T\simeq 4\times 10^6$.

Next, we briefly discuss the general-relativistic
Thomas-Fermi theory
\cite{bil}
for a self-gravitating gas,
consisting of $N$ fermions of mass $m$
in equilibrium at a temperature $T$,
enclosed in
a sphere of radius $R$.
We denote by
$p$, $\rho$, and $n$
the  pressure,
energy density, and  particle number density
of the gas,
respectively.
The metric generated by the mass distribution
is  static, spherically symmetric, and asymptotically
flat, i.e.,
\begin{equation}
ds^2=\xi^2 dt^2 -(1-2{\cal{M}}/r)^{-1} dr^2 -
     r^2(d\theta^2+\sin \theta d\phi^2).
\label{eq13}
\end{equation}
For numerical convenience,
we introduce the  parameter
\begin{equation}
\alpha=\frac{\mu}{T}
\label{eq86}
\end{equation}
and the substitution
\begin{equation}
\xi=\frac{\mu}{m}(\varphi+1)^{-1/2},
\label{eq87}
\end{equation}
where $\mu$ is the chemical potential associated with
the conserved
particle number $N$.
Using this, the equation
of state for a self-gravitating ideal gas may be represented
in a parametric form \cite{ehl}:
\begin{equation}
n=\frac{1}{\pi^2}
               \int^{\infty}_{0} dy
\frac{y^2}
{1+\exp \{[(y^2+1)^{1/2}/(\varphi+1)^{1/2}-1]\alpha\} },
\label{eq83}
\end{equation}
\begin{equation}
\rho=\frac{1}{\pi^2}
               \int^{\infty}_{0}dy\,
\frac{y^2(y^2+1)^{1/2}}
{1+\exp \{[(y^2+1)^{1/2}/(\varphi+1)^{1/2}-1]\alpha\} },
\label{eq84}
\end{equation}
\begin{equation}
p=\frac{1}{3\pi^2}
               \int^{\infty}_{0}dy\,\frac{y^4
(y^2+1)^{-1/2}}
{1+\exp \{[(y^2+1)^{1/2}/(\varphi+1)^{1/2}-1]\alpha\} }.
\label{eq85}
\end{equation}
We have chosen appropriate length and mass  scales
$a$ and $b$,  respectively, such that
\begin{equation}
a=b=
\sqrt{\frac{2}{g}}\,
\frac{1}{m^2},
\label{eq81}
\end{equation}
or,  restoring $\hbar$, $c$, and $G$,
we have
\begin{equation}
a=
\sqrt{\frac{2}{g}} \,
\frac{\hbar M_{\rm{Pl}}}{c m^2}
=1.0798 \times 10^{10}\,
\sqrt{\frac{2}{g}} \,
\left(
\frac{15{\rm keV}}{m}
\right)^2
{\rm km},
\end{equation}
\begin{equation}
b=
\sqrt{\frac{2}{g}} \,
\frac{M_{\rm{Pl}}^3}{m^2}
=0.7251 \times 10^{10}\,
\sqrt{\frac{2}{g}} \,
\left(
\frac{15{\rm keV}}{m}
\right)^2
M_{\odot} \, .
\end{equation}
Here, $M_{\rm{Pl}}=\sqrt{\hbar c/G}$
denotes the Planck mass and $g$
 the combined spin-degeneracy factor
 of neutral fermions and antifermions,
 i.e., g=2 or 4 for Majorana or Dirac fermions, respectively.
In this way, the fermion mass, the degeneracy factor,  and
the chemical potential are eliminated from the
equation of state.

Einstein's field equations for the metric (\ref{eq13})
are given by
\begin{equation}
\frac{d\varphi}{dr} =
-2(\varphi+1)\frac{{\cal M}+4\pi r^3 p}{r(r-2{\cal{M}})} \, ,
\label{eq88}
\end{equation}
\begin{equation}
\label{eq89}
\frac{d{\cal{M}}}{dr}=4\pi r^2 \rho.
\end{equation}
To these two equations we add
\begin{equation}
\frac{d{\cal N}}{dr}=4\pi r^2 (1-2{\cal{M}}/r)^{-1/2} n ,
\label{eq93}
\end{equation}
imposing
 the particle-number constraint as
a condition at the boundary
\begin{equation}
{\cal N}(R)=N.
\label{eq94}
\end{equation}
Equations (\ref{eq88})-(\ref{eq93})
should be integrated using
the boundary conditions at the origin:
\begin{equation}
\varphi(0)=\varphi_0 > -1
\, ; \;\;\;\;\;
{\cal{M}}(0)=0
\, ; \;\;\;\;\;
{\cal{N}}(0)=0.
\label{eq90}
\end{equation}

It is useful to introduce the degeneracy
parameter
$\eta=\alpha \varphi/2$,
which,
in the Newtonian limit,
approaches
$\eta_{\rm nr}=(\mu_{\rm nr} -V)/T$.
Here, we have introduced
the nonrelativistic chemical potential
$\mu_{\rm nr}=\mu-m$,
with $\mu_{\rm nr}\ll m$,
and the approximation
$\xi=e^{V/m}\simeq 1+V/m$, with $V$ being the Newtonian potential.
As $\varphi$ is monotonously decreasing with increasing
 $r$, the strongest degeneracy is obtained at the center
 with $\eta_0=\alpha\varphi_0/2$.
The parameter $\eta_0$,
uniquely related to the central
density and pressure,
will eventually be fixed
by the requirement (\ref{eq94}).
For $r\geq R$, the function $\varphi$ yields
the usual empty-space Schwarzschild
solution
\begin{equation}
\varphi(r)=\frac{\mu^2}{m^2}
\left(1-\frac{2 M}{r}\right)^{-1}-1\, ,
\label{eq91}
\end{equation}
  with
\begin{equation}
M={\cal M}(R)=\int_0^R dr 4\pi r^2 \rho(r) .
\label{eq92}
\end{equation}
Given the temperature $T$, the set of self-consistency
equations (\ref{eq83})-(\ref{eq93}), with the boundary
conditions (\ref{eq94})-(\ref{eq92})
defines the general-relativistic Thomas-Fermi
equation.

The numerical procedure is now straightforward.
For a fixed, arbitrarily chosen
$\alpha$,
 we first  integrate
Eqs.\
 (\ref{eq88}) and (\ref{eq89})
 numerically
on the interval $[0,R]$ and find
solutions
 for various central values
$\eta_0$.
Integrating (\ref{eq93}) simultaneously,
we obtain
${\cal N}(R)$ as a function of $\eta_0$.
We then select the value of $\eta_0$
for which
${\cal N}(R)=N$.
The chemical potential $\mu$
corresponding to
this particular solution
 is given by
(\ref{eq91}).
If we now eliminate $\mu$
using (\ref{eq86}), we finally get the
parametric dependence on temperature
through $\alpha$.

\begin{figure}[t]
\centering
\epsfig{file=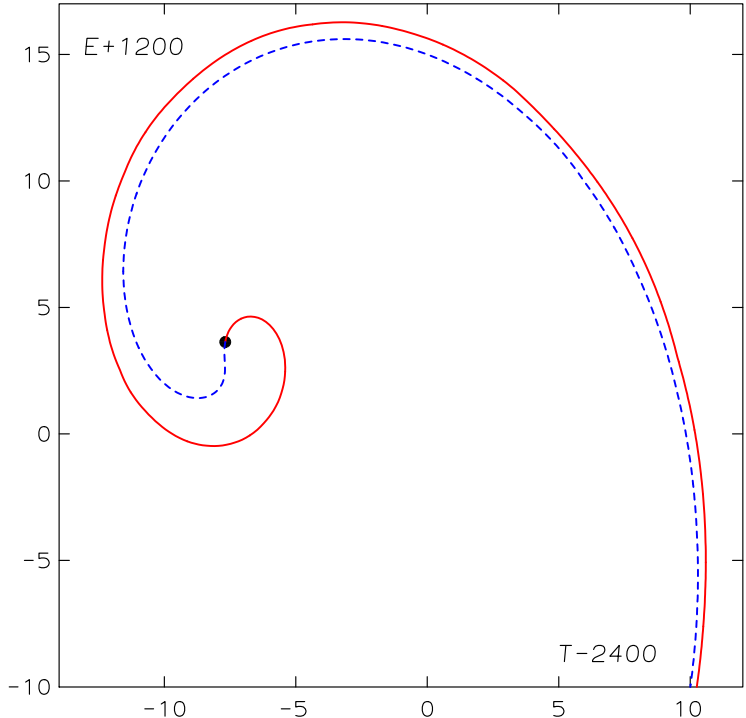,width=10cm}
\caption{
Energy (shifted by $12\times 10^{-8}m$) versus
temperature (shifted by $-24\times 10^{-8}m$), both in
units of $10^{-10}m$, for fixed $N=2\times 10^{12} M_{\odot}/m$
}
\label{fig1}
\end{figure}

The quantities
$N$, $T$, and, $R$ are free parameters in our model
and their
range and choice are dictated
by physics.
At $T=0$
the number of fermions $N$  is
 restricted by the OV limit
$N_{\rm OV}=2.89
\times 10^{9}\,
\sqrt{2/g}
(15\,{\rm keV}/m)^2
M_{\odot}/m $.
However, at nonzero temperature, stable solutions exist
with $N>N_{\rm OV}$, depending on temperature
and radius.
In the following,  $N$ is required to be
of the order $2\times 10^{12} M_{\odot}/m$,
so that for any $m$, the total mass
is close to the estimated mass of the halo
\cite{wilk}.
As we have demonstrated,
the expected particle mass-temperature ratio of the halo
is given by $\alpha\simeq m/T=4\times 10^4$.
 The halo radius
$R$ is in principle unlimited; in practice, however,
it should not exceed half the average intergalactic distance.
It is known that an
 isothermal configuration has no natural boundary,
in contrast to the degenerate
 case of zero temperature,  where
for given $N$ (up to the OV limit)
the radius $R$ is naturally fixed by the condition of
vanishing pressure and density.
At nonzero  temperature,
with $R$ being unbounded, our gas would occupy
the entire space, and
fixing $N$ would make
$p$ and $\rho$
vanish everywhere.
Conversely,
if we do not fix $N$ and integrate the
equations  on the interval $[0,\infty)$,
both $M$ and $N$ will
diverge at infinity for $T>0$.
Thus, one is forced to introduce a cutoff.
In an isothermal model of a similar kind
\cite{cha},
the  cutoff was set
at the radius $R$, where the energy density
was by about six orders of magnitude smaller than the central value.
Our choice of
$R=200$ kpc
is based on the estimated size of the Galactic halo.

The only remaining free parameters of our model are the fermion
mass $m$ and the degeneracy factor $g$,
which always appear in the combination
$m^4g$.
We fix these parameters at
$m=15$ keV and $g=2$, and justify this choice {\em a posteriori}.

\begin{figure}[t]
\centering
\epsfig{file=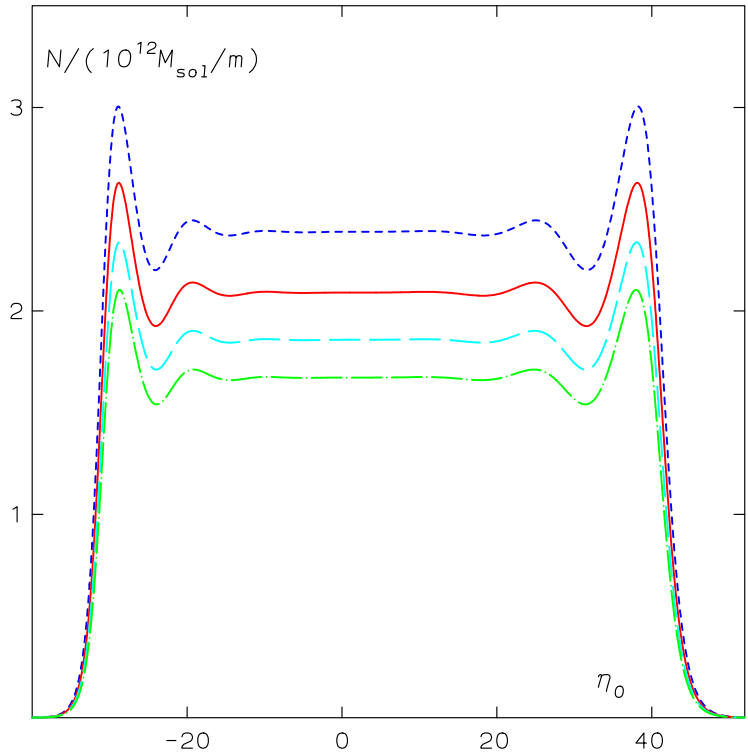,width=10cm}
\caption{
Number of particles  versus central degeneracy
parameter
for $m/T= 4\times 10^6$
(solid),
$3.5\times 10^6$
(short dashs),
$4.5\times 10^6$
(long dashs), and
$5\times 10^6$
 (dot-dashed line).
}
\label{fig2}
\end{figure}

We  now present the results of the calculations for  fixed
particle number and  temperatures near the
point of gravothermal collapse.
In Fig.\ \ref{fig1} the energy per particle
defined as $E=M/N-m$
is plotted
as a function of temperature for fixed
$N=2\times 10^{12} M_{\odot}/m$.
The plot looks very much like that of
a canonical Maxwell-Boltzmann  ensemble \cite{bin},
with one important difference:
in the Maxwell-Boltzmann case, the curve  continues to
spiral inwards {\em ad infinitum} approaching the point
of the singular isothermal sphere,
that is
characterized by an infinite central
density.
In the Fermi-Dirac case, the spiral consists of two, almost
identical curves.
The inwards winding of the spiral begins
for some  negative central degeneracy and stops at the point
$T=2.3923 \times 10^{-7} m$,
$E=-1.1964 \times 10^{-7} m$,
where $\eta_0$ becomes zero.
This part of the curve, which basically depicts the behavior
of a nondegenerate gas, we call
the {\em Maxwell-Boltzmann branch}.
By increasing the central degeneracy parameter
further to positive values,
the spiral begins to unwind outwards very close to
the inwards winding curve.
The outwards winding curve
will eventually depart from the
Maxwell-Boltzmann branch for temperatures $T \simgt 10^{-3} m$.
Further increase of the central degeneracy parameter brings us to
a region  where general-relativistic effects become
important.
The curve will exhibit another spiral
for temperatures and energies of the order of a few $10^{-3}m$
approaching the limiting temperature  $T_{\infty}= 2.4 \times 10^{-3}m$
and energy
$E_{\infty}= 3.6 \times 10^{-3}m$, with both  the central degeneracy
parameter
and the central density approaching infinite values.
It is remarkable that
gravitationally stable configurations
with  arbitrary large central degeneracy parameters
exist
at finite temperature
even though the total
mass exceeds the OV limit by several
orders of magnitude.

The results of the numerical integration of Eqs.\
 (\ref{eq88}) and (\ref{eq89}),
 without restricting $N$, are presented
in Fig.\ \ref{fig2}, where we plot the  particle number $N$
as a function of the central
degeneracy parameter
$\eta_0$ for several
values of $\alpha$ close to $4\times 10^6$.
For fixed $N$, there is a range of $\alpha$,
where the Thomas-Fermi equation has multiple solutions.
For example, for $N=2\times 10^{12}$ and
$\alpha=4\times 10^6$ six solutions are found,
which are  denoted by
(1), (2), (3), (3'), (2'), and (1'),
corresponding to the values $\eta_0 =$
 -30.528,
 -25.354,
 -22.390,
  29.284,
  33.380, and
  40.479, respectively.
\begin{figure}[t]
\centering
\epsfig{file=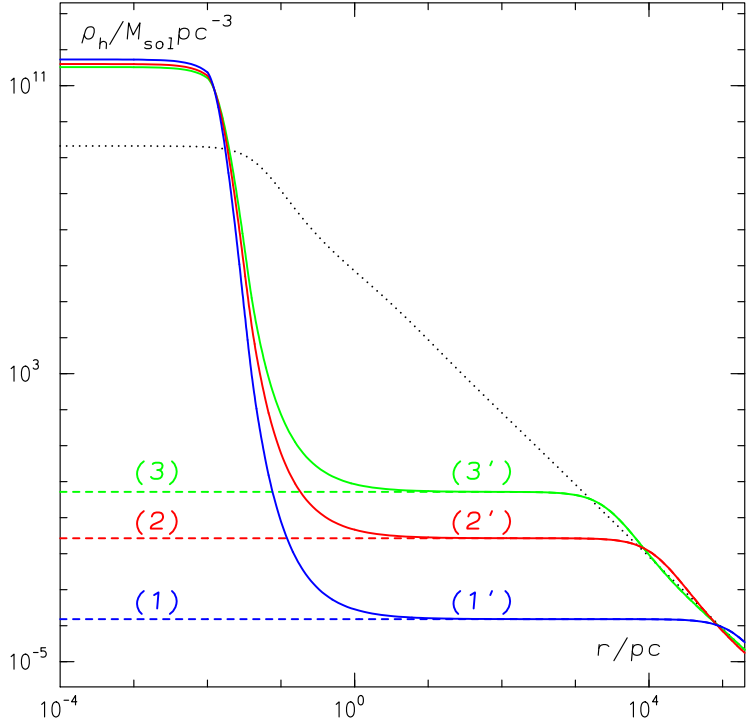,width=10cm}
\caption{
The density profile of the halo
for a central degeneracy parameter $\eta_0=0$ (dotted line) and for
the six $\eta_0$-values discussed in the text.
Configurations with negative $\eta_0$
((1)-(3)) are depicted by the dashed
and those with positive $\eta_0$
((1')-(3')) by the solid line.
}
\label{fig3}
\end{figure}
In Figs.\ \ref{fig3}
and \ref{fig4} we plot the corresponding density profiles and enclosed
masses, respectively.
For negative central value $\eta_0$,
for which the degeneracy parameter is negative everywhere,
the system behaves basically as a
Maxwell-Boltzmann isothermal sphere.
Positive values of the central degeneracy parameter $\eta_0$
are characterized by a pronounced central core
of mass of about $2.5 \times 10^6 M_{\odot}$
within a radius of about 20 mpc.
The presence of this core is obviously due to
the degeneracy pressure of the
Fermi-Dirac statistics.
A similar structure was obtained in
collisionless stellar systems modeled as
a nonrelativistic Fermi gas
\cite{chav}.
\begin{figure}[t]
\centering
\epsfig{file=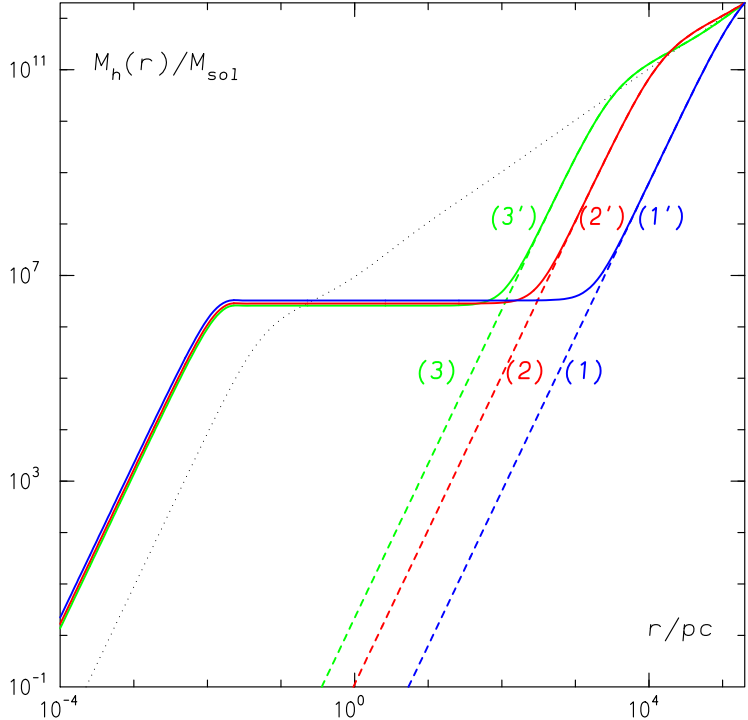,width=10cm}
\caption{
Mass of the halo $M_{\rm h}(r)$ enclosed within a radius $r$
for various central degeneracy parameters
$\eta_0$ as in Fig.\ \protect\ref{fig3}.
}
\label{fig4}
\end{figure}

\begin{figure}[t]
\centering
\epsfig{file=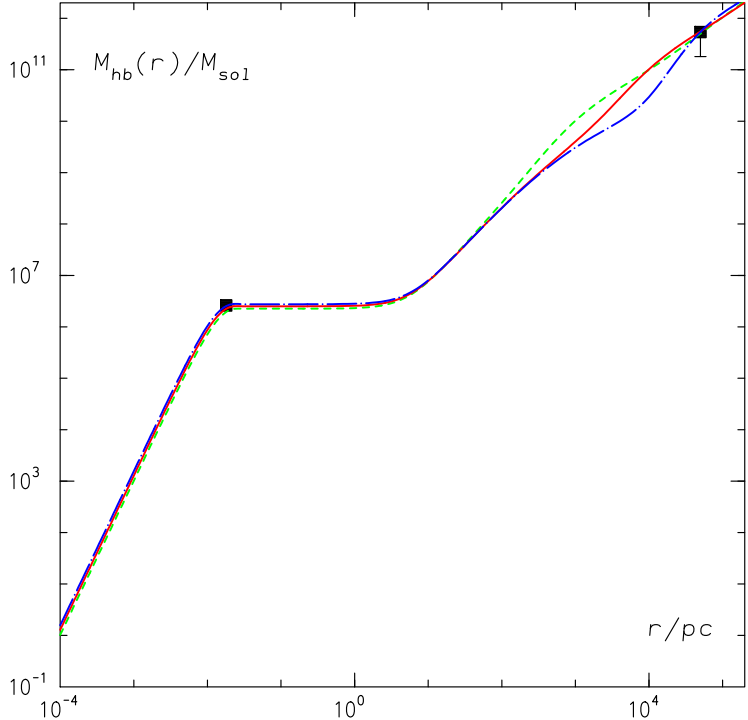,width=10cm}
\caption{
Enclosed mass of halo plus bulge versus
radius for $\eta_0$ =
   24 (dashed),
  28 (solid),
and
    32  (dot-dashed line).
}
\label{fig5}
\end{figure}

Figs.\ \ref{fig3}
and \ref{fig4} show two important features.
First,
a galactic halo at a given temperature $T$
may or may not have a central core,
depending on whether  the central degeneracy parameter $\eta_0$
is positive or negative.
As the potential is nearly harmonic
up to about 1 to 10 kpc for negative
$\eta_0$, this may favor the formation of a barred
galaxy.
Second,
the closer to zero $\eta_0$ is,
the smaller the radius is at which the
$r^{-2}$  asymptotic behavior of the density begins.
The flattening of the Galactic rotation  curve
begins in the range  $1 \simlt r/{\rm kpc} \simlt 10$,
hence the solution (3') most likely describes the
Galaxy's halo.
This may be verified by calculating the rotational
curves in our model.
We know already from our estimate (\ref{eq005})
that our model
yields the correct asymptotic circular velocity of
220 km/s.
In order to make a more realistic comparison
with the observed Galactic rotation curve,
we must include
two additional matter components: the bulge and
the disk.
The bulge is modeled as a spherically symmetric matter distribution
of the form
\cite{you}
\begin{equation}
\rho_{\rm b}(s)=\frac{e^{-hs}}{2s^3}
\int_0^{\infty} du
\frac{e^{-hsu}}{[(u+1)^8-1]^{1/2}} \, ,
\label{eq006}
\end{equation}
where $s=(r/r_0)^{1/4}$, $r_0$ is the effective radius of the bulge
and $h$ is  a parameter.
We adopt  $r_0=2.67$ kpc
and $h$ yielding the bulge mass
$M_{\rm b}= 1.5 \times 10^{10} M_{\odot}$
\cite{suc}.
In Fig.\ \ref{fig5} the mass
of halo and bulge enclosed within
a given radius is plotted for various $\eta_0$.
Here, the gravitational backreaction  of the bulge on
the fermionic halo has been taken into account.
\noindent
The data points, indicated  by squares, are
the  mass
$M_{\rm c}=2.6 \times 10^6 M_{\odot}$ within
18 mpc, estimated from the motion of the stars
near Sgr A$^*$ \cite{eck},
and the mass
$M_{50}=5.4^{+0.2}_{-3.6}\times 10^{11}$
within 50 kpc,
 estimated from
the motions
of satellite galaxies and globular clusters
\cite{wilk}.
Variation of the central degeneracy parameter
$\eta_0$ between 24 and 32 does not change
the essential halo features.

\begin{figure}[t]
\centering
\epsfig{file=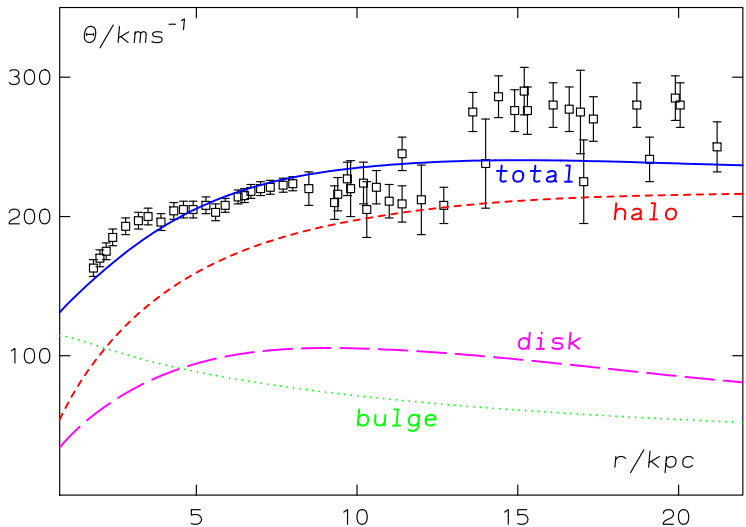,width=10cm}
\caption{
Fit to the rotation curve of the Galaxy.
The data points are from \protect\cite{oll}
for $R_0=8.5$ kpc and $\Theta_0=220$ km/s.
}
\label{fig6}
\end{figure}

In Fig.\ \ref{fig6} we plot
the circular velocity components of
the halo, the bulge, and the disk.
The contribution of the disk
 is modeled  as \cite{per}
\begin{equation}
\Theta_{\rm d}(r)^2=
\Theta_{\rm d}(r_{\rm o})^2
\frac{1.97 (r/r_{\rm o})^{1.22}}{
[(r/r_{\rm o})^2+0.78^2]^{1.43}} \, ,
\label{eq007}
\end{equation}
where we have taken
$r_{\rm o}=13.5$ kpc and
$\Theta_{\rm d}=100$ km/s.
Here it is assumed for simplicity that
the disk does not influence the mass distribution
of the bulge and the halo.
Choosing the central degeneracy
$\eta_0=28$ for the halo, the data
by Merrifield and Olling \cite{oll} are reasonably well
fitted.

We now turn to the discussion of
our choice of the fermion mass $m=15$ keV
for the degeneracy factor $g=2$.
To that end,
we  investigate how  the mass of the
central object,
i.e., the mass $M_{\rm c}$ within 18 mpc,
depends on $m$ in the interval
5 to 25 keV,
for various
$\eta_0$.
We find that $m\simeq15$ keV always gives the maximal value of
$M_{\rm c}$
ranging between 1.7 and 2.3 $\times 10^6 M_{\odot}$
for $\eta_0$ between 20 and 28.
Hence, with  $m\simeq 15$ keV  we get the
value closest to the mass of the central object
$M_{\rm c}$
estimated from the motion of the stars
near Sgr A$^*$ \cite{eck}.

\section{Conclusions}
In summary,
using the Thomas-Fermi
theory, we have shown that
a  weakly interacting
fermionic gas  at finite temperature
yields  a mass distribution that
successfully describes both the center and the halo
of the Galaxy.
For a fermion mass
$m \simeq 15$ keV,
a reasonable fit to the rotation
curve is achieved with the
temperature $T = 3.75$ meV and
the central degeneracy parameter
$\eta_0=28$.
With the same parameters,
we obtain
the mass
$M_{50} = 5.04\times 10^{11} M_{\odot}$
and
$M_{200} = 2.04\times 10^{12} M_{\odot}$
within 50 and 200 kpc, respectively.
These values agree quite well with the mass estimates
based on the motions
of satellite galaxies and globular clusters
\cite{wilk}.
Moreover, the mass
of $M_{\rm c} \simeq 2.27\times 10^6 M_{\odot}$,
enclosed within 18 mpc,
agrees reasonably  well
with the observations of
motion of stars near
the compact dark object at
the center of the Galaxy.

\subsection*{Acknowledgement}

We thank P.\ Salucci for valuable discussions and comments.
We are grateful to
M.R.\ Merrifield and
R.P.\ Olling for sending us the
Galactic rotation curve.
This
research is in part supported by the Foundation of Fundamental
Research (FFR) grant number PHY99-01241 and the Research Committee of
the University of Cape Town.  The work of N.B.\ is supported in part by
the Ministry of Science and Technology of the Republic of Croatia
under Contract No.\ 0098002.

\end{document}